\documentclass[12pt]{article}

\usepackage[utf8]{inputenc}
\usepackage{ngerman,a4}

\usepackage{graphicx,epsfig}
\usepackage{color}

\usepackage{amssymb,amsmath,amsfonts}
\usepackage{bbold}

\usepackage{cite}

\def\etc{\textit{etc}}

\def\rest{\upharpoonright}

\def\cA{{\cal A}}

\def\cH{{\cal H}}

\def\cK{{\cal K}}
\def\cL{{\cal L}}
\def\cM{{\cal M}}

\def\cP{{\cal P}}

\def\cR{{\cal R}}
\def\cS{{\cal S}}

\def\NN{{\mathbb N}}

\def\RR{{\mathbb R}}

\newcommand{\scirc}{\raisebox{1.2pt}{$\, \scriptscriptstyle \circ \, $}}

\title{\vspace*{-20mm}
\bf Masselose Teilchen und Zeitpfeil \\
in der relativistischen Quantenfeldtheorie}

\author{Detlev Buchholz \\[10mm]
Institut f\"ur Theoretische Physik und Courant Forschungszentrum \\
``Strukturen höherer Ordnung in der Mathematik'', 
Universit\"at G\"ottingen, \\ 37077 G\"ottingen, Deutschland}

\date{} 

\begin{document}

\maketitle

\begin{abstract} \noindent
Die übliche Interpretation der physikalischen Zustände 
in Quantenfeldtheorien vom Typ der Quanten\-elektrodynamik 
bereitet Schwierigkeiten aufgrund des Auftretens 
masseloser Teilchen, wie dem Photon. In Stoßprozessen
geladener Teilchen werden unendlich viele dieser 
masselosen Teilchen erzeugt, die zu einer enormen Vielfalt 
von Zuständen führen zwischen denen 
theoretisch keine Interferenzen möglich sind (sie gehören
zu verschiedenen Superauswahlsektoren); experimentell
lassen sich diese Zustände jedoch nicht unterscheiden. Dieser
scheinbare Widerspruch löst sich auf wenn man berücksichtigt, 
dass reale Experimente stets in zukunftsgerichteten
Gebieten der Raumzeit (Lichtkegeln) durchgeführt werden. Versäumte
Messungen in der Vergangenheit lassen sich aufgrund
der Richtung des Zeitpfeils nicht nachholen. Die Theorie
kann sich daher auf die Beschreibung und Interpretation
solcher Lichtkegeldaten beschränken. In diesem 
Artikel wird erläutert, wie dieser Gedanke im allgemeinen 
Rahmen der Quantenfeldtheorie mathematisiert wird 
und zu einem konsistenten physikalischen Bild führt. Zustände 
gleicher Ladung aber inkohärenter masseloser Anteile 
lassen sich bei Einschränkung auf die Observablen in 
einem gegebenen Lichtkegel zu Ladungsklassen zusammenfassen;
jede dieser Klassen vereinigt eine Vielzahl von 
Superauswahlsektoren. Es zeigt sich, dass 
zu jeder Ladungsklasse eine konjugierte Klasse existiert,
die Zustände der entgegengesetzten Ladung 
(Antimaterie) beschreibt.
Alle Zustände in einer Klasse und der entsprechenden
konjugierten Klasse genügen der gleichen (Bose oder Fermi)
Statistik. Obwohl auf Lichtkegeln nur die 
Halbgruppe zukunftsgerichteter Zeittranslationen wirkt, 
lässt sich sowohl die relativistische Kovarianz als auch 
die energetische Stabilität (Positivität der Energie)
der Zustände jeder Ladungsklasse etablieren. Der
Formalismus ist somit geeignet, die Eigenschaften 
der Zustände in Quantenfeldtheorien mit langreichweitigen 
Kräften in physikalisch sinnvoller Weise zu beschreiben.
\end{abstract}

\section{Einleitung}

Das Verständnis der Struktur des physikalischen 
Zustandsraumes in Quantenfeldtheorien mit langreichweitigen 
Kräften, prominentestes Beispiel ist die Quantenelektrodynamik,  
ist ein sehr altes Problem. Seine unterschiedlichen Aspekte
wurden in zahlreichen Arbeiten untersucht, siehe zum 
Beispiel die entsprechenden Kapitel in den 
Monographien \cite{Ha,St,Str} und dort 
angegebene Referenzen. Allen diesen Untersuchungen 
ist gemeinsam, dass die physikalische Raumzeit als 
unendlich ausgedehnter Minkowskiraum modelliert wird,
in dem die quantisierte 
Materie und die von masselosen Teilchen 
erzeugte Strahlung beschrieben wird (die Gravitation
bleibt außer Acht).  
Geht man von diesem idealisierten Bild aus, so ergibt sich
als Konsequenz, dass man beliebig langwellige Strahlung
bzw.\ masselose Teilchen beliebig kleiner Energie 
behandeln muss, obwohl diese in realistischen Experimenten
nicht nachgewiesen werden können. Dies führt sowohl bei der 
Konstruktion der Theorie als auch bei ihrer Interpretation
zu Schwierigkeiten, die zuweilen als Infrarotkatastrophe 
bezeichnet werden. Man hat zwar gelernt, wie man diese
Schwierigkeiten im Rahmen der Theorie bei einigen 
Fragestellungen umgehen kann, etwa bei der Berechnung von 
``inklusiven'' Wirkungsquerschnitten von Stoßprozessen.
Dort summiert man über unendlich viele unbeobachtbare 
{niederenergetische} 
masselose Teilchen, die bei der 
Wechselwirkung geladener Teilchen unvermeidlich erzeugt werden. Doch 
ist diese Methode, die aufgrund der Überidealisierung der 
experimentellen Situation eingehandelten Schwierigkeiten 
durch \textit{ad hoc} Rezepte 
wieder zu beseitigen, konzeptionell nicht befriedigend. 
Zudem bleiben grundlegende Fragen, etwa nach der Sektorstruktur
des physikalischen Zustandsraumes, nach den Wurzeln der 
Teilchenstatistik oder der Existenz von Antimaterie bei
diesem Zugang unbeantwortet. Bei Abwesenheit von
langreichweitigen Kräften konnten diese Fragen dagegen  
vor langem umfassend beantwortet werden \cite{Ha}. 

Ein ganz neuer Zugang zu diesem Problemkreis, bei dem
die raumzeitlichen Einschränkungen bei realen Experimenten
von vornherein im Rahmen der Theorie berücksichtigt werden, wurde in 
einer kürzlich erschienenen Arbeit vorgeschlagen \cite{BuRo}.
Die gewonnen Einsichten sollen hier erläutert werden ohne
auf technische Details einzugehen. Vorausgesetzt werden beim 
Leser lediglich Kenntnisse von Grundbegriffen der Quantentheorie 
und der speziellen Relativitätstheorie. 

\section{Grundlagen}

Es sei zunächst daran erinnert, dass die Ensembles eines physikalischen
Systems durch normierte Vektoren (Zustandsvektoren) $\Phi$
eines Hilbertraumes $\cH$ beschrieben werden, die Messgrößen 
 (Observablen) wirken auf diesem Raum durch  
lineare hermitesche Operatoren $A = A^*$. Operatoren lassen sich 
addieren, per Komposition multiplizieren und mit komplexen Zahlen 
skalieren, sie erzeugen durch diese Operationen eine Algebra $\cA$. 
Die Theorie macht Vorhersagen über Mittelwerte der
Messgrößen in vorgegebenen Ensembles, deren 
Schwankungsquadrate \etc. Sie werden 
beschrieben durch Erwartungswertfunktionale $\varphi$ auf der 
Observablenalgebra $\cA$. Diese Funktionale erhält man aus den   
Hilbertraumzuständen $\Phi$ gemäß der fundamentalen Formel 
$$ \varphi(A) = \langle \Phi, A \Phi \rangle \, ,
\quad A \in \cA \, , $$
wobei, wie üblich, die eckige Klammer das Skalarprodukt zwischen
Hilbertraumvektoren beschreibt. 
Es ist hier wichtig, dass man umgekehrt aus jedem solchen 
(linearen und positiven) Erwartungswertfunktional --
wir sprechen im Folgenden kurz von Zuständen --
einen Hilbertraum und eine konkrete Darstellung der 
Observablen auf diesem Raum rekonstruieren kann, so 
dass obige Relation gilt (sogenannte GNS--Konstruktion \cite{Ha}). 
Dies ist insbesondere dann von Bedeutung, wenn man 
Ensembles behandeln möchte, die nicht von vornherein 
durch Vektoren im vorgegebenen Hilbertraum beschrieben
werden können (zum Beispiel ladungstragende Zustände
wenn man von neutralen Zuständen ausgeht). 
Sie lassen sich durch geeignete Grenzwerte der 
zur Verfügung stehenden Funktionale darstellen. 
Man kann so den gegebenen Hilbertraum 
``verlassen'' und zu einer konsistenten Beschreibung
der gewünschten Ensembles unter Beibehaltung der 
Observablen kommen. Genauer gesagt, man erhält aus den 
Limesfunktionalen mittels der GNS--Konstruktion eine 
neue Darstellung der gegebenen Observablenalgebra  
auf einem Hilbertraum. In ihm gelten nach wie vor die ursprünglichen
algebraischen Relationen zwischen den Observablen 
(man denke an Vertauschungsrelationen,
Feldgleichungen \etc). 

In der relativistischen Quantenfeldtheorie benutzt man 
zweckmäßigerweise das Heisenbergbild. Das heißt, die 
Symmetrietransformationen im zugrundeliegenden  
Minkowskiraum $\cM$, wie die Verschiebung des 
Zeitpunktes und Ortes von Messungen 
oder Drehungen der Messgeräte und der Übergang von einem 
Inertialsystem zu einem anderen 
wirken auf die entsprechende Observablen in~$\cA$.
Für jede derartige Poincar\'e--Transformation $\lambda$ wird 
diese Wirkung beschrieben durch eine 
lineare, multiplikative und symmetrische 
Abbildung $\alpha_\lambda$ von $\cA$
auf sich (Automorphismus). Man hat ebenfalls Informationen über 
die raumzeitlichen Lokalisierungseigenschaften der 
Observablen.
Messungen in einem gegebenen Raumzeitgebiet $\cR \subset \cM$
werden beschrieben durch Observablen in entsprechenden 
Unteralgebren $\cA(\cR) \subset \cA$. Da diese Messungen 
auch in jedem größeren Gebiet ausgeführt werden können, gilt
aus Konsistenzgründen $\cA(\cR_1) \subset \cA(\cR_2)$
falls $\cR_1 \subset \cR_2$. Weiterhin ist die Wirkung der 
Symmetrietransformationen $\lambda$ in $\cM$ konsistent 
mit der Zuordnung von Observablen zu Gebieten, 
d.h.\ Observablen in $\cA(\cR)$ werden 
durch die Wirkung von $\alpha_\lambda$ in die Algebra $\cA(\lambda
\cR)$ des transformierten Gebietes $\lambda \cR$ überführt,
in Formeln $\alpha_\lambda \, \cA(\cR) = \cA(\lambda \cR)$. 

Das Einsteinsche Kausalitätsprinzip,
also die Aussage, dass sich physikalische Effekte nicht mit
Überlichtgeschwindigkeit ausbreiten (dies gilt auch für die 
Quanteneffekte von Messungen), lässt sich in diesem 
Rahmen ebenfalls in sehr einfacher Weise formulieren. Dazu sei daran
erinnert, dass Messungen, die sich gegenseitig nicht  
beeinflussen, durch Observablen beschrieben werden, deren
Produkt kommutativ ist. Gemäß
Kausalitätsprinzip müssen daher Paare von  
Observablen in raumartig getrennten (nicht durch kausale Weltlinien
verbindbaren) Gebieten $\cR_1, \cR_2$ miteinander kommutieren,
d.h.\ $A_1 A_2 = A_2 A_1$ für alle $A_1 \in \cA(\cR_1)$.
$A_2 \in \cA(\cR_2)$; man bezeichnet 
dies als Lokalitätsbedingung. Jede 
physikalische Theorie, die den Grundprinzipien 
der {Quantentheorie} und der speziellen Relativitätstheorie genügt, 
erfüllt diese Bedingungen. Es ist bemerkenswert, dass der 
so abgesteckte allgemeine Rahmen vollständig ausreicht, um 
die hier interessierenden konzeptionellen Fragen zu diskutieren. 

Geht man von der Vorstellung aus, dass der zugrunde gelegte 
Hilbertraum $\cH$ die Zustände sämtlicher elementarer Systeme
einer Theorie beschreibt (z.B.\ alle Zustände endlicher Energie)
so stellt sich die Frage nach dessen Sektorstruktur. Dabei trägt 
man der Tatsache Rechnung, dass in der Quantenfeldtheorie 
das Superpositionsprinzip nur eingeschränkt gilt: 
Zustandsvektoren unterschiedlicher Gesamtladung lassen sich 
zwar auf dem Papier addieren, ihre relative Phase spielt 
jedoch experimentell keine Rolle, sie sind nicht kohärenzfähig.
Die Familien kohärenzfähiger Zuständen in $\cH$ nennt man
(Superauswahl) Sektoren. Sie kann man dadurch charakterisieren,
dass man von jedem gegebenen Zustand in einem Sektor zu jedem
anderen Zustand im gleichen Sektor durch die Quanteneffekte von Messungen 
gelangen kann, die Observablen operieren transitiv  
auf den Sektoren. Verschiedene Sektoren lassen sich 
durch klassische Observablen unterscheiden, die man als Grenzwerte 
von geeigneten Folgen von Observablen in $\cA$ erhält. Sie 
vertauschen mit allen Elementen von $\cA$ und haben in jedem 
Sektor einen scharfen (nicht statistisch fluktuierenden) Wert,
die entsprechende Superauswahlladung des Sektors. 

Prominente Beispiele solcher Superauswahlladungen sind die 
elektrische Ladung und die sogenannte Univalenz, mit der man 
bosonische und fermionische Sektoren unterscheiden kann. Es
zeigt sich jedoch, dass es in Quantenfeldtheorien mit langreichweitigen 
Kräften, die durch masselose Teilchen vermittelt werden, eine 
ungeheure Vielzahl von weiteren Sektoren gibt. Sie unterscheiden sich 
durch facettenreiche Infrarotwolken unendlich vieler masseloser 
Teilchen, die bei Stoßprozessen geladener Teilchen erzeugt werden.   
Die theoretische Beschreibung und Klassifikation aller dieser 
Infrarotsektoren ist mathematisch ein hoffnungsloses Problem. 
Zum Glück spielt sie experimentell keine Rolle. Denn es ist 
praktisch unmöglich unendliche Konfigurationen masseloser 
Teilchen mit beliebig kleiner Energie scharf zu unterscheiden. 
Bei Anwendungen der Theorie trägt man dieser Tatsache meist 
dadurch Rechnung, dass man \textit{ad hoc} 
spezielle Sektoren auswählt und in Erwartungswerten 
über alle unbeobachtbaren 
\mbox{niederenergetischen} masselosen Teilchen 
summiert. Dadurch kann man zwar eine Reihe von Schwierigkeiten
umgehen, die Methode ist jedoch aus zwei Gründen 
konzeptionell unbefriedigend. Zum einen erfordert die 
Unterscheidung zwischen hoch-- und niederenergetischen
Teilchen die Wahl eines Inertialsystems, man bricht damit 
die relativistische Invarianz der Theorie. Zum anderen zerstören 
die Quanteneffekte der lokalisierten Observablen diese 
Unterscheidung wegen der \mbox{Heisenbergschen} Unschärferelation.
Jedes Ensemble niederenergetischer 
Teilchen enthält nach einer lokalen Messung auch Anteile 
beliebig hoher Energie. Diese Tatsache erschwert
die Diskussion von Konsequenzen des Einsteinschen 
Kausalitätsprinzips bei diesem Zugang. 

Die kürzliche Lösung dieses Problems beruht auf der Einsicht, dass 
der Zeitpfeil bereits bei der Interpretation
der mikroskopischen Theorie berücksichtigt werden muss. Um 
Missverständnissen vorzubeugen: der Zeitpfeil soll
hier nicht erklärt werden, er wird lediglich als empirische 
\mbox{Tatsache} in den theoretischen Rahmen eingebaut. Es ist ein
Faktum, dass es unmöglich ist, versäumte Messungen in 
der Vergangenheit heute oder in der Zukunft nachzuholen. 
Man kann also nicht ausschließen, dass gewisse Daten,
die man in der Vergangenheit hätte bestimmen können,
unwiederbringlich verloren sind. Anders gesagt, 
die Theorie kann sich darauf beschränken, die Ergebnisse 
von Messungen in 
Raumzeitgebieten zu beschreiben und zu erklären, die 
in realistischen Experimenten (zumindest prinzipiell) 
zugänglich sind. Dies sind vorwärts gerichtete Lichtkegel $V$
mit Spitze in einem beliebig gewählten 
Raumzeitpunkt $a$ in der Vergangenheit; der 
Rand des Kegels wird von den von der Spitze ausgehenden Lichtstrahlen 
gebildet. 
\begin{figure}[h] 
 \hspace*{60mm}
\epsfig{file=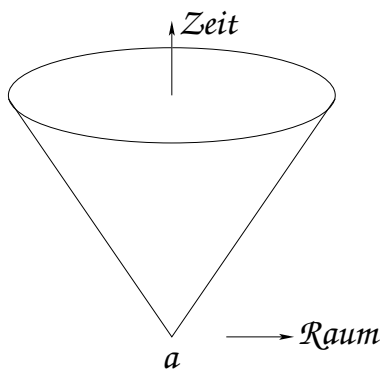,height=48mm}
\caption{\ Lichtkegel $V$ über dem Raumzeitpunkt $a$}
\label{zeitpfeil} 
\end{figure}
Wo die Spitze des Kegels liegt spielt praktisch keine Rolle. 
Man kann z.B.\ den Geburtstag und Geburtsort von Aristoteles wählen,
der das Wort Physik ($\phi \upsilon \sigma \iota \kappa \eta$)
eingeführt hat; wir alle wissen davon 
und befinden uns also innerhalb des entsprechenden Kegels. 
Eine andere Wahl wäre z.B.\ der Tag
und Ort, an dem die Finanzierung eines geplanten 
Experiments genehmigt wurde. Wichtig ist nur, dass in 
der Vergangenheit des gewählten Punktes $a$ keine experimentellen
Daten berücksichtigt werden müssen bzw.\ können. 
Natürlich erstrecken sich reale Experimente 
nur über einen bestimmten Zeitraum. Doch 
könnten im Prinzip zukünftige Generationen ein Experiment
bis in alle Ewigkeit weiterführen. Lichtkegel sind also 
die maximalen Raumzeitgebiete in denen Experimente möglich sind.  
  
Im Rahmen der Theorie werden die Observablen in einem
gegebenen Lichtkegel $V$ durch Elemente der entsprechenden Unteralgebra
$\cA(V) \subset \cA$ beschrieben. Geht man davon aus, dass
nur diese Observablen in einem Experiment zur Verfügung stehen,
so kann man bei gegebenem Zustand~$\varphi$ eines Systems nur 
die entsprechenden Erwartungswerte $\varphi(A)$, $A \in \cA(V)$ bestimmen.
Man erhält also nur partielle Informationen über den 
globalen Zustand $\varphi$ und bezeichnet dessen Einschränkung
$\varphi \rest \cA(V)$ daher als partiellen Zustand. 
Es stellt sich dann die Frage, welche 
Information man aus solchen partiellen Zuständen
extrahieren kann. 

Bei genauerer Analyse zeigt sich, dass in Theorien, in 
denen es ausschließlich massive Teilchen gibt (also in einer 
hypothetischen Welt), die partiellen Zustände bereits die vollständige
Information über den globalen Zustand enthalten, egal wie man
den Lichtkegel $V$ wählt, d.h.\ wann und wo man mit den Messungen 
begonnen hat. Dies kann man auch heuristisch verstehen wenn man bedenkt, dass
die Weltlinien massiver Teilchen, die sich ja mit weniger 
als Lichtgeschwindigkeit bewegen, irgendwann einmal in den 
Lichtkegel eintreten müssen. Aus den Daten in~$V$ 
lassen sich daher die globalen Daten mit Hilfe der 
Theorie rekonstruieren. 

\begin{figure}[h] 
 \hspace*{60mm}
\epsfig{file=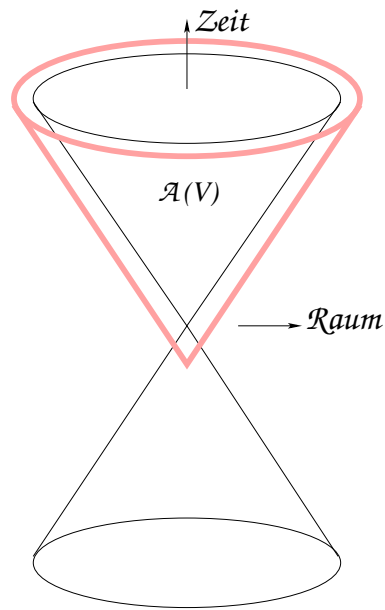,height=80mm}
\caption{In der Vergangenheit erzeugte masselose Teilchen 
entgehen Messungen in $V$}
\label{huygens}
\end{figure}

Sobald masselose Teilchen in der 
Theorie auftreten (``Es werde Licht'') ändert sich die Situation
jedoch drastisch. Werden in der 
Vergangenheit des  gegebenen Lichtkegels $V$ durch 
Wechselwirkungsprozesse masselose Teilchen erzeugt, die 
danach ungehindert durch die Raumzeit propagieren, so 
können diese $V$ nicht erreichen und dort nachgewiesen 
werden, denn sie bewegen sich stets 
mit Lichtgeschwindigkeit (Huygensches Prinzip). Aus den 
\mbox{partiellen} Zuständen $\varphi \rest \cA(V)$ lassen sich 
diese auslaufenden masselosen Anteile daher nicht rekonstruieren. Man 
erhält somit in Experimenten nur eingeschränkte Informationen über 
die Beiträge masseloser Teilchen in den Zuständen.
Wie wir noch sehen werden, ist diese
Tatsache dafür verantwortlich, dass man die unterschiedlichen
Infrarotwolken in den partiellen Zuständen nicht 
scharf unterscheiden kann. Die Zustände in verschiedenen
Infrarotsektoren stimmen bei Restriktion auf 
Lichtkegel überein und man kann auf diese Weise Klassen von 
physikalisch äquivalenten (weil ununterscheidbaren)  
Zuständen bilden. Von Bedeutung ist dabei, dass 
die masselosen Teilchen in den  Zuständen nicht mehr 
in einen energetisch weichen und einen harten
Anteil aufgeteilt werden (mit den erwähnten Problemen). 
Stattdessen wird der masselose Teilcheninhalt getrennt 
in einen marginalen (nicht notwendig niederenergetischen) 
Teil, der in $V$ keine Beiträge liefert 
und einen essentiellen Teil, der Messungen 
in $V$ zugänglich ist. Diese Aufspaltung ist mit der 
relativistischen Invarianz der Theorie und dem 
Einsteinschen Kausalitätsprinzip verträglich und
bildet den Ausgangspunkt der Untersuchungen in \cite{BuRo},
die wir im Folgenden erläuteren wollen.

\section{Das Vakuum}

Sowohl bei der Konstruktion von Theorien als auch bei deren 
Interpretation spielt der Zustand des Vakuums eine ausgezeichnete 
Rolle. Dies liegt daran, dass man eine ganze Reihe von 
Eigenschaften dieses speziellen 
Zustandes nennen kann, ohne die Theorie konstruiert zu 
haben: Der Vakuumzustand ist für alle inertialen Beobachter
gleich, das heißt Messergebnisse in diesem Zustand 
ändern sich nicht, wenn man die Positionen der Messapparate  
durch Poincar\'e--Transformationen ändert.
Ferner ist das Vakuum energielos, d.h. für alle inertialen
Beobachter ist es der Zustand niedrigster Energie (Grundzustand).

Da wir nur Messungen in Lichtkegeln betrachten wollen, 
müssen wir bei der mathematischen Umsetzung dieser 
Charakterisierung beachten, dass die von den 
Poincar\'e--Transformationen 
$\lambda$ erzeugte Poincar\'e--Gruppe
$\cP$ keine Symmetriegruppe der Kegel ist; denn kein 
Lichtkegel $V$ ist invariant unter beliebigen derartigen
Transformationen. Die Untergruppe $\cL$ der 
Drehungen und Geschwindigkeitstransformationen 
bei festgehaltener Spitze des Kegels (Lorentz--Gruppe)
lässt $V$ zwar invariant, doch ändert jede raumzeitliche 
Translation die Lage von $V$ im Minkowski--Raum.
Es gibt jedoch Translationen, bei denen  man $V$ 
zumindest nicht verlässt. Dies sind zukunftsgerichtete 
zeitartige (oder auch lichtartige) Verschiebungen. 
Inertiale Beobachter in $V$ müssen wenig tun, 
um derartige Verschiebungen ihrer Messgeräte zu 
bewerkstelligen: sie müssen nur warten, der 
Zeitpfeil tut das Übrige. Die Familie der
Poincar\'e--Transformationen, die $V$ auf bzw.\ in sich abbilden, 
erzeugt eine Semigruppe $\cS \subset \cP$ 
(``semi'' deshalb, weil die Elemente von $\cS$ sich zwar 
komponieren lassen, es gibt jedoch kein Inverses, da sich
der Zeitpfeil nicht umkehren lässt).

Nach diesen Vorbemerkungen können wir nun obige 
formlose Beschreibung des Vakuums in unserem Formalismus 
präzisieren. 
Ein partieller Zustand $\omega \rest \cA(V)$ wird als
Vakuum in $V$ gedeutet wenn 
$$
\omega(\alpha_\lambda(A)) = \omega(A) 
\quad \mbox{für alle} \quad A \in \cA(V), \ \lambda \in \cS \, .
$$
Man ergänzt diese Relation noch durch einige physikalisch 
sinnvolle Bedingungen an die Natur der Korrelationen zwischen 
Paaren von Observablen in diesem Zustand, siehe \cite{BuRo}.
Mit Hilfe der weiter oben erwähnten GNS--Konstruktion 
kann man dann einen Hilbertraum $\cH$ konstruieren, auf dem 
die Observablen in $\cA(V)$ operieren, sowie einen 
Zustandsvektor $\Omega \in \cH$ angeben, so dass 
$\omega(A) = \langle \Omega, A \Omega \rangle$ für alle 
$A \in \cA(V)$. Die Vektoren in $\cH$ beschreiben 
die durch Quanteneffekte von Messungen erzeugten 
lokalen Anregungen des Vakuums. Der interessante Punkt ist 
nun, dass die oben erwähnten charakteristischen 
Eigenschaften des partiellen Vakuumzustandes bereits 
ausreichen, um auf $\cH$ 
eine Darstellung der \textit{vollen} Poincar\'e Gruppe $\cP$
durch unitäre Operatoren $U(\lambda)$, $\lambda \in \cP$
zu konstruieren \cite{BuRo}. Für ihre Einschränkung 
auf $\cS$  gelten die Relationen 
$$ 
U(\lambda) A U(\lambda)^{-1} = \alpha_\lambda(A) \, , \quad
U(\lambda) \Omega = \Omega \, , \quad \lambda \in \cS \, .
$$
Man kann also auch im Fall der auf Lichtkegel eingeschränkten
Messungen alle Symmetrietransformationen durch unitäre 
Operatoren beschreiben, die sich aus den entsprechenden \mbox{Daten}
in \mbox{kanonischer} Weise rekonstruieren lassen. 
\begin{figure}[h] 
\hspace*{40mm}
\epsfig{file=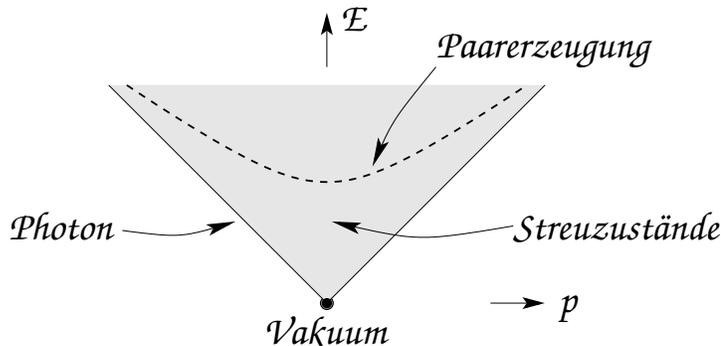,height=45mm}
\caption{Energie--Impulsspektrum in der Quantenelektrodynamik}
\label{spektrum} 
\end{figure}
Darüber hinaus 
haben die Generatoren der resultierenden Raumzeittranslationen, die wie
üblich als Energie--Impulsoperatoren gedeutet werden, 
die erwarteten spektralen Eigenschaften (Positivität der 
Energie für alle inertialen Beobachter). Sie lassen sich jedoch
nicht in der üblichen Weise als Observablen deuten,  
da in ihnen die Fluktuationen des Energieinhalts in $V$ infolge  
von in der Vergangenheit abgestrahlten masselosen Teilchen 
nur pauschal berücksichtigt werden. Die Situation ist ähnlich wie 
in der Quantenstatistischen Mechanik, wo 
die Unkenntnis der Mikrozustände thermischer Systeme 
in die Liouvilleoperatoren statistisch eingeht. 
Ein charakteristisches Energie--Impulsspektrum, wie man es zum
Beispiel für die lokalen Anregungen des 
Vakuums in der \mbox{Quantenelektrodynamik} erwartet, 
ist in Figur \ref{spektrum} dargestellt. Eingezeichnet  
sind die Zustände, die zu den jeweiligen Energie--Impulswerten
beitragen: Die Spitze des Kegels
entspricht dem Vakuum, der Rand  
den Energie--Impulswerten  eines einzelnen 
Photons. Die Spektralwerte in der Nähe des Randes gehören 
zu Streuzuständen von Photonen (Delbrück--Streuung).
Photonen genügend hoher Energie können bei Stoßprozessen 
Elektron--Positron Paare erzeugen, \etc.

\section{Ladungsklassen}

Auch wenn das Vakuum für die Konstruktion und Interpretation
der Theorie von Bedeutung ist, so sind doch vor allem seine 
lokalen Anregungen und die daraus resultierenden 
ladungstragenden Zustände von physikalischem 
Interesse. Bevor wir uns der Konstruktion und
Analyse dieser Zustände zuwenden, müssen wir zunächst 
besser verstehen, weshalb die
bei der Interpretation der Theorie im Minkowskiraum auftretenden
Infrarotprobleme bei unserem Zugang verschwinden.

\begin{figure}[h] 
 \hspace*{62mm}
\epsfig{file=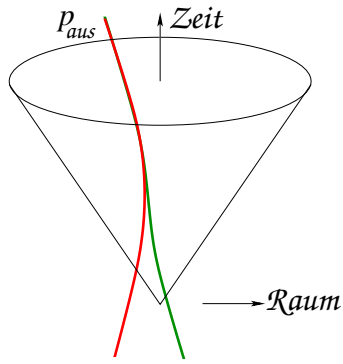,height=50mm}
\caption{Die Sektoren der Infrarotwolken sind in Lichtkegeln 
nicht unterscheidbar}
\label{fock} 
\end{figure}

Gemäß der Maxwellschen Elektrodynamik erzeugen 
beschleunigte Punktladungen Strahlung. Im Rahmen der 
Quantenfeldtheorie kann man diese Strahlung durch kohärente
Zustände von Photonen beschreiben. Es stellt sich bei
genauerer Analyse heraus, dass für Punktladungen 
mit verschiedenem einlaufenden und auslaufenden Impuls
$p_{\mbox{\tiny \it ein}} \neq p_{\mbox{\tiny \it aus}}$ diese Zustände mit 
Sicherheit unendlich viele 
niederenergetische Photonen enthalten (sie können 
nicht durch Vektorzustände im Fockraum der Photonen beschrieben werden).
\begin{figure}[b] 
 \hspace*{37mm}
\epsfig{file=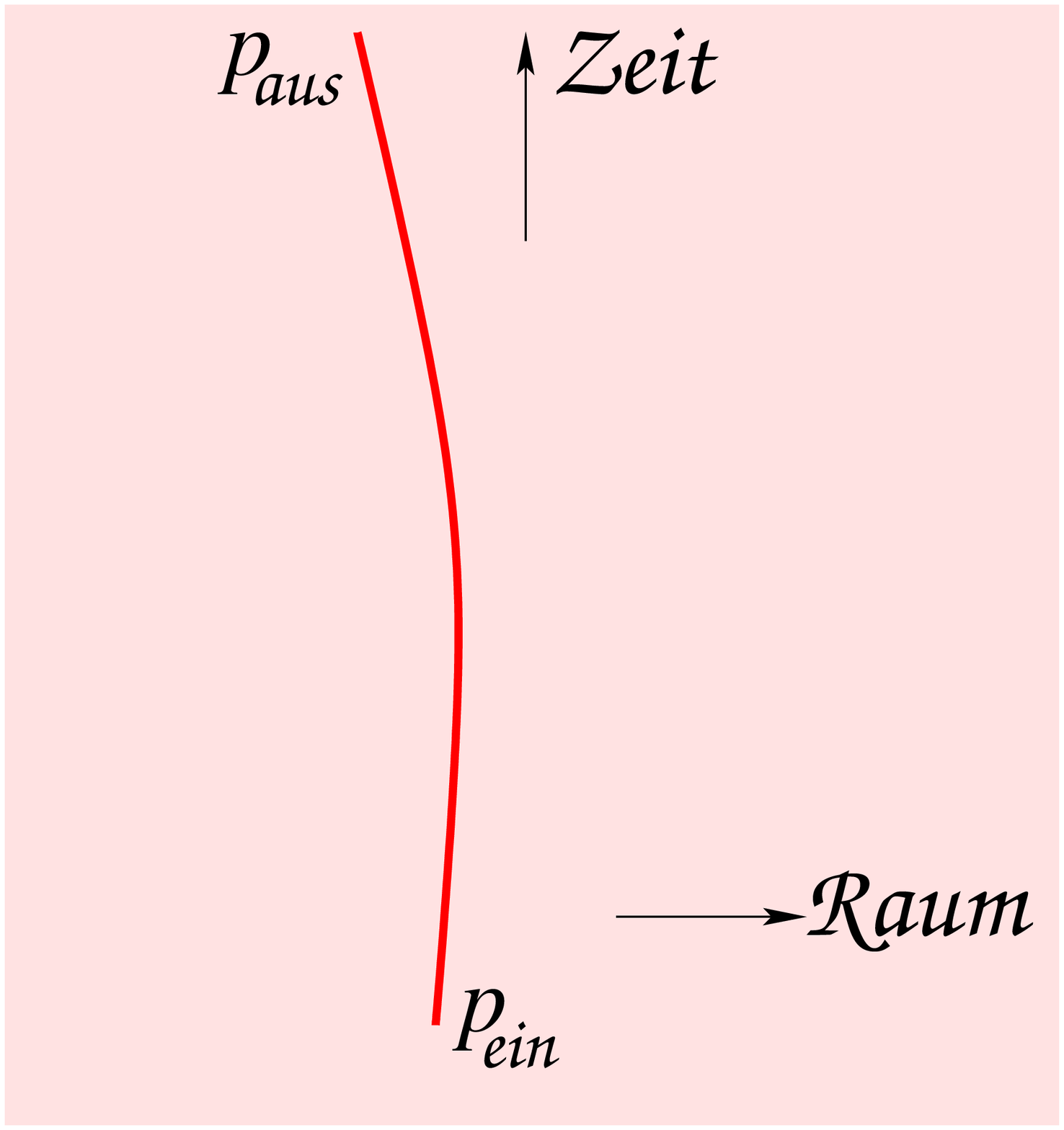,width=40mm,height=50mm}
 \hspace*{10mm}
\epsfig{file=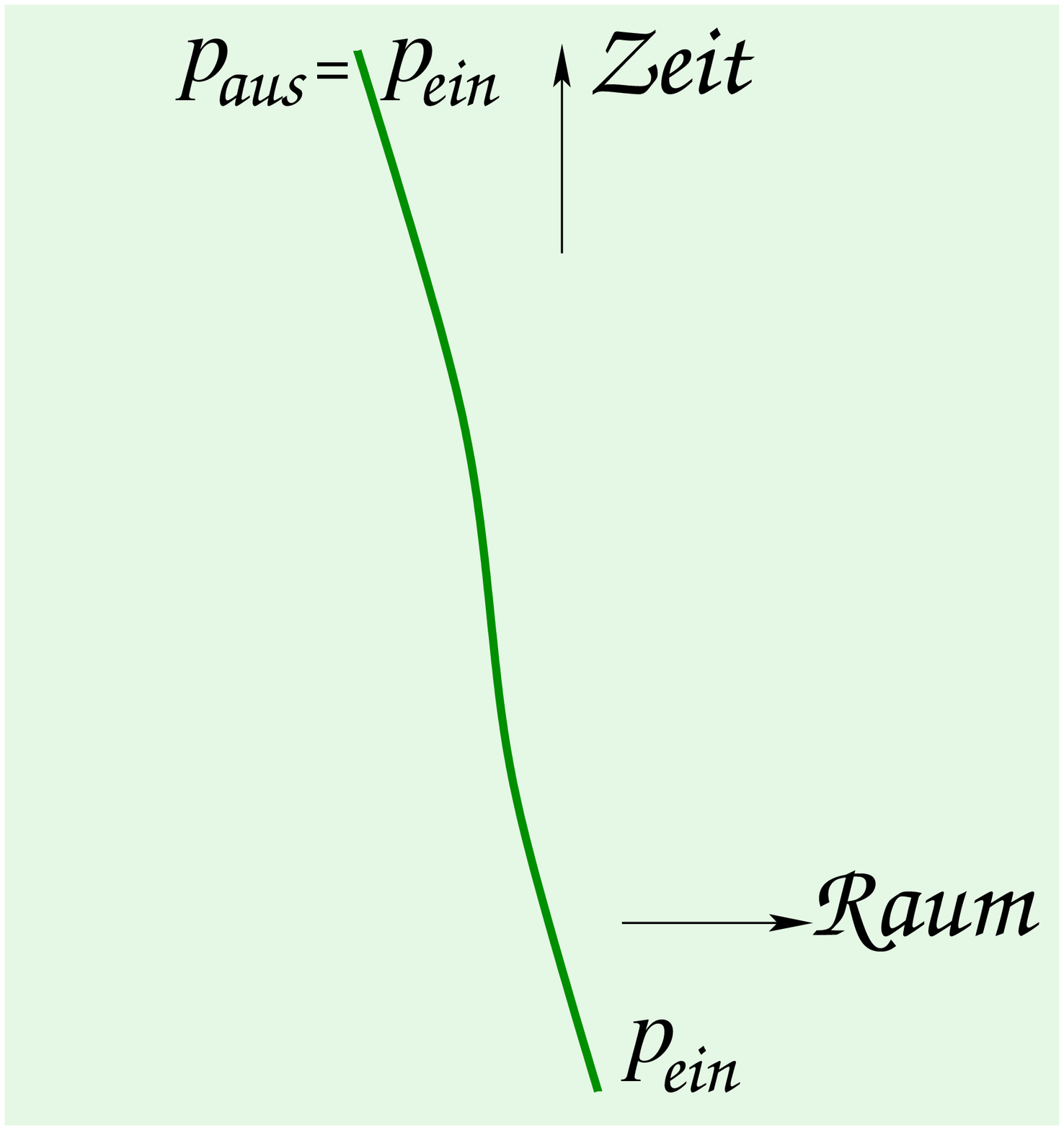,width=40mm,height=50mm}
\caption{Beschleunigte Ladungen erzeugen global unterscheidbare
Infrarotwolken (Sektoren)}
\label{stroeme}   
\end{figure}
Man spricht daher, wie bereits mehrfach erwähnt, von Infrarotwolken.
Es zeigt sich ferner, dass diese Infrarotwolken \mbox{empfindlich} 
von den präzisen Werten der asymptotischen Impulse abhängen. Es 
gibt daher eine ungeheure Vielfalt von solchen global unterscheidbaren  
(im Sinne von Superauswahlsektoren) 
Infrarotwolken. Doch gibt es einen interessanten Spezialfall:
Stimmen die beiden asymptotischen Impulse der Punktladung exakt 
überein, so erhält man kohärente Zustände im Fockraum der Photonen,
d.h.\ mit endlicher Teilchenzahl. Diese Tatsache bleibt
meist unbeachtet, da es ja extrem unwahrscheinlich ist, dass
in realen Prozessen ein solches Ereignis auftritt. Doch spielt 
gerade dieser Spezialfall bei unseren Überlegungen eine wichtige
Rolle. Stellt man sich nämlich die Frage, ob sich die 
Photonenwolken der beiden in Fig.~\ref{stroeme} 
angedeuteten Prozesse durch Messungen in einem 
Lichtkegel~$V$ scharf \mbox{unterscheiden} lassen, 
so verdeutlicht Fig.~\ref{fock} dass dies nicht so sein sollte. 
Denn durch Messung an den Photonen in $V$ 
wird sich nicht mit
Sicherheit feststellen lassen, welche einlaufenden 
Impulse zwei Punktladungen mit dem gleichen auslaufenden
Impuls in ferner Vergangenheit einmal hatten; 
durch Wechselwirkungsprozesse in der \mbox{Vergangenheit} 
und dabei abgestrahlte marginale Photonen wird
diese in den Superauswahlsektoren enthaltene globale 
Information völlig verwischt. Die Messergebnisse 
werden daher konsistent sein mit denen, die man in 
Zuständen mit übereinstimmenden einlaufenden und auslaufenden  
Impulsen  erhalten würde. Anders gesagt, globale 
Infrarotwolken sollten sich bei Messungen in $V$ 
nicht von Zuständen im Fockraum der Photonen unterscheiden
lassen. Tatsächlich lässt sich dies im 
Rahmen der Quantenfeldtheorie beweisen \cite{Bu}.

\begin{figure}[h] 
 \hspace*{60mm}
\epsfig{file=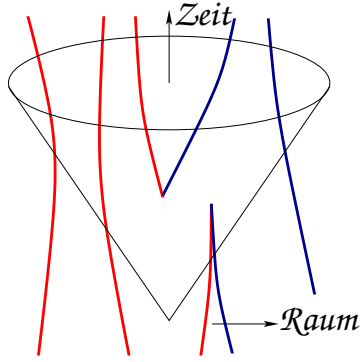,height=48mm}
\caption{Die globale Ladung lässt sich in Lichtkegeln bestimmen}
\label{ladung}
\end{figure}

Während die Infrarotsektoren sich also bei Messungen in Lichtkegeln 
nicht unterscheiden lassen, kann der Gesamtwert der von massiven
Teilchen in einem Zustand  
getragenen Ladungen in Lichtkegeln ermittelt 
werden. Dies veranschaulicht Fig. \ref{ladung}, in der die
Weltlinien von massiven geladenen Teilchen eingezeichnet sind,
die sich natürlich alle mit weniger als Lichtgeschwindigkeit bewegen. 
Massive Teilchen entgegengesetzter Ladung können in 
Stoßprozessen vernichtet werden (wobei z.B.\ hochenergetische 
Photonen entstehen) oder aber durch solche Prozesse in Paaren
entgegengesetzter Ladung erzeugt werden. Diese Ereignisse 
ändern die \mbox{Gesamtladung} eines Zustands nicht. Man kann daher
aus der Analyse des massiven Teilcheninhalts 
in \mbox{Lichtkegeln} die von diesen Teilchen 
getragene Gesamtladung bestimmen. Diese wichtige 
Einsicht lässt sich im Rahmen der Quantenfeldtheorie 
präzise formulieren und begründen \cite{Bu}.

In Analogie zu den Superauswahlsektoren im Minkowskiraum $\cM$
kann man daher im Rahmen der Theorie die
Zustände mit gleicher Gesamtladung und auf gegebenem Lichtkegel
$V$ ununterscheidbaren Infrarotwolken in Ladungsklassen einteilen.
Dabei besteht jede Ladungsklasse aus Zuständen, deren Einschränkungen 
auf $\cA(V)$ durch die Quanteneffekte 
physikalischer Operationen in $V$ ineinander überführt werden
können. Wir erinnern daran, dass Operationen in der Quantentheorie
durch die Wirkung von unitären Operatoren auf Zustandsvektoren
beschrieben werden: Stört man die Dynamik des Systems 
für einige Zeit durch einen in $V$ lokalisierten Operator, so
kann man den Effekt dieser Störung im Wechselwirkungsbild
berechnen. Nach Abschaltung der Störung erhält man 
einen unitären Operator $W \in \cA(V)$, der die Änderung
der Zustände aufgrund dieser Operation beschreibt. 
Dies führt zu folgender Definition,
die wir hier in etwas vereinfachter Form wiedergeben
und auch so benutzen werden. Die präzise Formulierung 
findet sich in \cite{BuRo}. 

\vspace*{2mm} 
\noindent \textbf{\textit{Ladungsklassen:}} \ 
Zwei reine (ungemischte) Zustände $\varphi_1$, $\varphi_2$ 
auf der Observablenalgebra $\cA$ gehören zur
gleichen Ladungsklasse, wenn es zu jedem gegebenem 
Lichtkegel $V$ einen entsprechenden unitären Operator 
$W \in \cA(V)$ gibt, so dass 
$$ \varphi_2 \rest \cA(V) = \varphi_1 \scirc \mbox{Ad} \, W \rest
\cA(V) \, ,$$ 
d.h.\ die partiellen Zustände werden durch die adjungierte 
Wirkung $\mbox{Ad} \, W = W \cdot W^{-1}$ von $W$ 
auf $\cA(V)$ ineinander überführt. 
Das Symbol $\scirc$ bezeichnet die Komposition von Abbildungen. 

\section{Ladungen  und Morphismen}

Geht man, so wie bei der Konstruktion von  
Quantenfeldtheorien üblich, vom 
Vakuumzustand aus, so kann man mit Hilfe der vorangegangenen 
Definition alle Zustände in der Ladungsklasse des Vakuums
charakterisieren, also alle neutralen Zustände, die keine 
globale Ladung tragen. Es stellt sich
dann die Frage, wie man im Rahmen der Theorie 
aus diesen neutralen Zuständen geladene Zustände erhält. 
Auch hier ergibt sich die Antwort aus physikalische Überlegungen. 

\begin{figure}[h] 
 \hspace*{60mm}
\epsfig{file=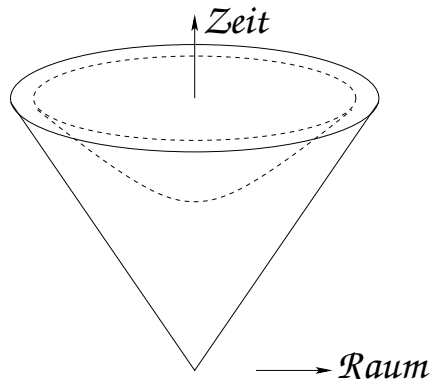,height=50mm}
\caption{Lichtkegel mit Zeitschale}
\label{zeitschale}
 \end{figure}

Wir wählen im Folgenden einen festen Lichtkegel $V$. Stellt man
sich vor, dass inertiale Beobachter, die in der Spitze des Kegels
mit unterschiedlichen Geschwindigkeiten in verschiedene Richtungen 
gestartet sind, alle zur gleichen Eigenzeit den Raumzeitpunkt, den
sie erreicht haben, markieren, so erhält man einen Hyperboloid
in $V$, den wir als
Zeitschale bezeichnen, siehe Fig.\ref{zeitschale}. 
Die asymptotisch lichtartigen Punkte auf einer solchen
Zeitschale, die man nur mit Lichtgeschwindigkeit erreichen könnte, 
liegen raumartig zu allen kompakten Gebieten im Inneren von $V$. 

\begin{figure}[h] 
 \hspace*{23mm}
\epsfig{file=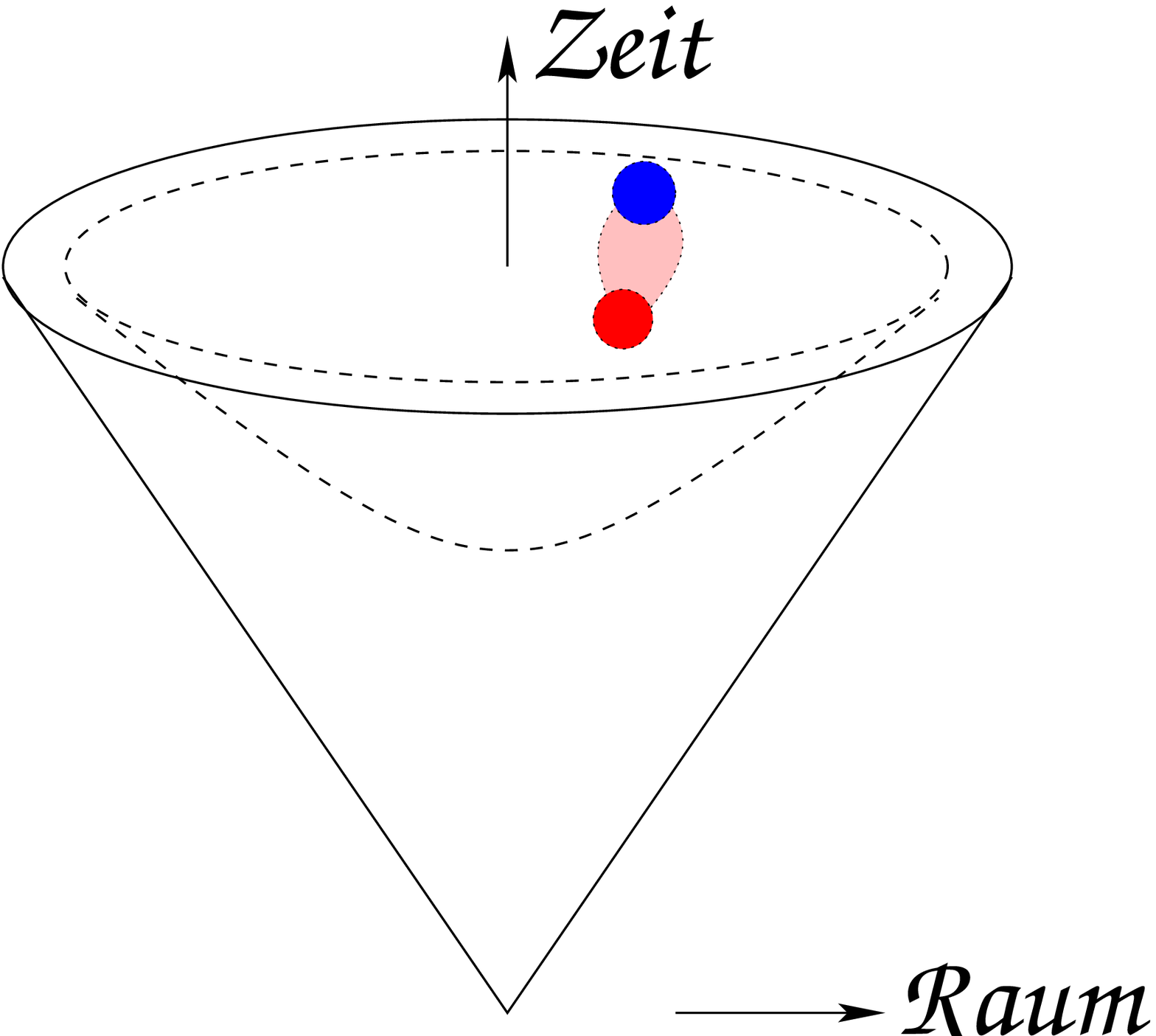,height=48mm}
\hspace*{13mm}
\epsfig{file=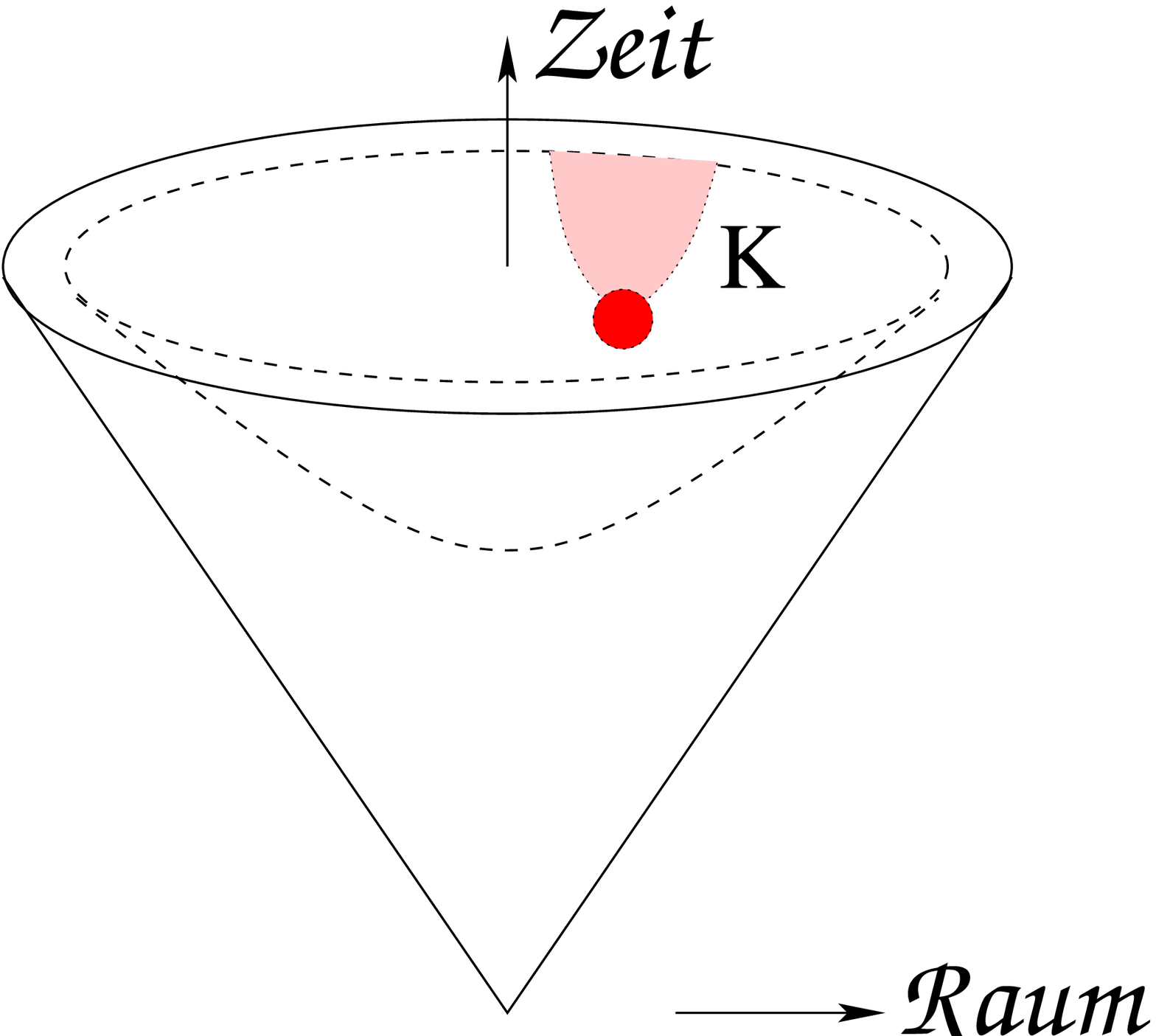,height=48mm}
\caption{Paarerzeugung und asymptotische Ladungserzeugung 
auf einer Zeitschale}
 \label{ladungen}
 \end{figure}

Durch lokale Operationen mit genügend großem Energieübertrag 
kann man nun auf einer gegebenen Zeitschale aus
dem Vakuum Paare von Teilchen mit entgegengesetzter Ladung erzeugen 
(Vakuumpolarisation), daneben entstehen Wolken von 
masselosen Teilchen. Um aus solchen global neutralen Paaren einen 
für Beobachter in $V$ effektiv ladungstragenden 
\mbox{Zustand} zu präparieren, muss die unerwünschte entgegengesetzte Ladung 
auf Lichtgeschwindigkeit gebracht werden. Sie befindet 
sich dann am asymptotisch lichtartigen Rand der Zeitschale und 
verschwindet dadurch aus dem kausalen Einflussgebiet von 
Beobachtern im Inneren von $V$. Dieser Limes würde in der 
Praxis unendlich viel Energie erfordern, die von der
kompensierenden Ladung weggetragen wird. Man betrachtet hier also, 
wie häufig in der Theoretischen Physik, eine Idealisierung, 
da sie es gestattet, die Eigenschaften der Ladungen  
in Reinkultur zu analysieren. 
Wie in Fig.~\ref{ladungen} 
angedeutet, kann die Erzeugung von Paaren in kompakten 
Gebieten erfolgen, die 
Erzeugung der geladenen Limeszustände erfordert dagegen unendlich 
ausgedehnte Gebiete, die sich zur Begrenzung der von den masselosen 
Teilchen abgestrahlten Energie asymptotisch weiten müssen. Sie haben 
daher die Form hyperbolischer Kegel $\rm K$. Es ist hier wichtig, 
dass die Richtung und der Öffnungswinkel dieser Kegel 
bei der Präparation von Zuständen mit gegebener Ladung frei gewählt 
werden kann. Wir bezeichnen die kausalen Abhängigkeitsgebiete 
(Abschlüsse) der Kegel $\rm K$ mit kalligrafischen 
Buchstaben $\cK$ und nennen sie kurz Hyperkegel.

Nach diesen Vorbemerkungen können wir nun 
beschreiben, wie man in der Theorie, ausgehend vom Vakuumzustand 
$\omega$, geladene Zustände erhält.
Dazu gibt man sich zunächst einen 
Hyperkegel $\cK \subset V$ vor, in dem die Präparation erfolgen soll.
Dann wählt man Folgen von in $\cK$ lokalisierten unitären Operatoren, 
$\{W_n \in \cA(\cK)\}_{n \in \NN}$,  die die Operation der 
Paarerzeugung beschreiben, wobei
für wachsendes $n$ die kompensierende Ladung immer weiter
in asymptotisch lichtartiger Richtung verschoben wird.
Bei geeigneter Wahl der Operatorfolge konvergieren die
resultierenden partiellen Zuständen 
in der Ladungsklasse des Vakuums,
$\{ \omega \scirc \mbox{Ad} \, W_n \rest \cA(V) \}_{n \in \NN}$, 
gegen einen partiellen Zustand $\varphi \rest \cA(V)$
aus der gewünschten Ladungsklasse. Die kompensierende 
Ladung ist in $\varphi \rest \cA(V)$ nicht mehr nachweisbar, sie
befindet sich im raumartigen Komplement der in $V$ zur
Verfügung stehenden Observablen.

Ausgehend von dieser 
physikalisch transparenten aber umständlichen Methode 
kann man zu einer bequemeren, aber etwas
abstrakteren Beschreibung der Ladungserzeugung übergehen.
Es zeigt sich nämlich, dass auch die Folgen der von den
unitären Operatoren induzierten Abbildungen 
$\{ \mbox{Ad} \, W_n \}_{n \in \NN}$ 
der Observablenalgebra $\cA(V)$ auf sich selbst
konvergieren, wobei wir hier nicht auf die 
in \cite{BuRo} diskutierten 
mathematischen Details eingehen können. Für die 
Grenzwerte führen wir die Notation ein 
$$ \sigma_\cK = \lim_{n \rightarrow \infty} \,  \mbox{Ad} \, W_n
\, .  
$$
Der Index $\cK$ erinnert daran, dass $\sigma_\cK$ 
Grenzwert von Operationen im Hyperkegel $\cK$ 
ist. Die so definierten Abbildungen $\sigma_\cK$ sind,
wie ihre Approximationen, verträglich mit der  
algebraischen Struktur von $\cA(V)$: sie sind linear, 
multiplikativ und symmetrisch; der Einfachheit halber nehmen 
wir ferner an, dass ihr Bildbereich ebenfalls in $\cA(V)$ liegt. 
Wir bezeichnen diese Abbildungen kurz als 
(in~$\cK$ lokalisierte) Morphismen. 

Die in der Monographie \cite{Ha} erläuterten 
tiefen Resultate über die Sektorstruktur 
von Quantenfeldtheorien massiver Teilchen basieren auf der Analyse 
solcher Morphismen. Auch im vorliegenden Kontext
von Theorien mit masselosen Teilchen erweisen sie sich als
nützliches analytisches Instrument. Denn in 
ihnen sind in mathematisch präziser Weise alle physikalischen 
Informationen enthalten, die wir in den vorangegangenen 
Abschnitten zusammengetragen haben. Wir fassen sie in der
folgenden Liste zusammen, in der einige technische Details    
etwas vereinfacht dargestellt sind. (Die  
präzise Formulierung findet sich in \cite{BuRo}.)  

\vspace*{4mm} 
\noindent \textbf{\textit{Lokalisierte Morphismen:}} \ 
Zu gegebener Ladungsklasse
und beliebigem Hyperkegel $\cK \subset V$ gibt es 
einen auf $\cA(V)$ definierten lokalisierten Morphismus $\sigma_\cK$ 
mit den folgenden Eigenschaften.
\begin{itemize}

\vspace*{-2mm}
\item[(a)] 
$\omega \scirc \sigma_\cK \rest \cA(V)$ stimmt mit einem partiellen 
Zustand aus der gegebenen Ladungsklasse überein.

\vspace*{-2mm}
\item[(b)] Für jedes Raumzeitgebiet $\cR \subset V$, 
das $\cK$ enthält, gilt $\sigma_\cK(\cA(\cR)) \subseteq
\cA(\cR)$. In den Fällen wo man Gleichheit hat, bezeichnet man
die Ladungsklasse und die entsprechenden Morphismen als 
``einfach''.

\vspace*{-2mm}
\item[(c)] 
Für jedes Raumzeitgebiet $\cR \subset V$,
das raumartig von $\cK$ getrennt ist, gilt
$\sigma_\cK \rest \cA(\cR) = \iota$, 
worin $\iota$ die identische (triviale) Abbildung bezeichnet.

\vspace*{-2mm}
\item[(d)] Zu jedem Paar von Morphismen $\sigma_{\cK_1}$, $\sigma_{\cK_2}$,
die mit der Ladungsklasse affiliiert sind, existieren  
unitäre Operatoren $W_{2 1} \in \cA(V)$, 
die die Morphismen verknüpfen, 
\mbox{$\mbox{Ad} W_{2 1} \scirc \sigma_{\cK_1} = \sigma_{\cK_2}$}. 
Man nennt die Operatoren $W_{2 1}$ Verknüpfungsoperatoren.
\end{itemize}

\noindent Punkt (a) dieser Liste wurde bereits ausführlich 
erläutert. Die Punkte (b) und (c) folgen aus den Eigenschaften
der approximierenden Operatoren $W_n$.
Im Fall von (b) gilt $W_n \in \cA(\cK) \subset \cA(\cR)$ 
und somit $\mbox{Ad} \, W_n (\cA(\cR)) \subseteq \cA(\cR)$ und 
im Fall (c) gilt aufgrund der
Vertauschbarkeit von in $\cR$ und $\cK$ lokalisierten 
Operatoren (Lokalitätsprinzip)  
$\mbox{Ad} \, W_n \rest \cA(\cR) = \iota$.
Diese Eigenschaften übertragen sich auf 
den Limes $\sigma_\cK$. 
In Punkt (d) kommt die Tatsache zum Ausdruck, dass die 
Superauswahlsektoren der bei der Erzeugung von Ladungen unvermeidlich 
entstehenden Infrarotwolken bei Messungen in $V$ nicht
unterschieden werden können. Die Zustände 
$\omega \scirc \sigma_{\cK_1} \rest \cA(V)$
und $\omega \scirc \sigma_{\cK_2} \rest \cA(V)$
gehören daher zur gleichen Ladungsklasse und können 
durch die adjungierte Wirkung unitärer Operatoren in $\cA(V)$ ineinander 
transformiert werden. Diese Operatoren verknüpfen auch die
lokalisierten Morphismen. 

Wir werden uns im
Folgenden auf die Diskussion der in (b) charakterisierten  
einfachen Ladungsklassen bzw.\ Morphismen beschränken. 
Zu ihnen gehört die uns hier besonders interessierende 
elektrische Ladung und die Univalenz.

\section{Ladungskonjugation und Statistik}

Mit Hilfe der lokalisierten Morphismen lässt sich 
die Struktur der Ladungsklassen in der Quantenfeldtheorie 
in systematischer Weise analysieren und klassifizieren. 
Dazu muss man voraussetzen, dass die Theorie alle  
in Hyperkegeln $\cK$ prinzipiell 
möglichen Messungen und Operationen 
beschreibt, d.h. dass sich die Algebren $\cA(\cK)$ 
nicht durch Hinzufügen weiterer Operatoren vergrößern lassen, 
ohne mit dem Lokalitätsprinzip in Konflikt zu geraten.
Sie sind also in diesem Sinne maximal. Diese 
physikalisch motivierte Forderung 
an die Theorie lässt sich mathematisch in Form einer 
Dualitätsbedingung ausdrücken \cite{BuRo}.

Wie bereits angedeutet, werden wir uns im Folgenden auf die Diskussion
der einfachen Ladungsklassen beschränken, 
ohne dies jedes Mal zu erwähnen. Der erste und
entscheidende Schritt bei der Analyse ist der Beweis, dass sich 
Ladungsklassen komponieren lassen. Anschaulich gesprochen 
entspricht diese Komposition der Addition des Ladungsinhaltes
von Zuständen. Genauer gesagt, man kann zeigen \cite{BuRo}, dass   
für gegebene Ladungsklassen $L_1$, $L_2$ von Zuständen und Lichtkegel $V$ 
die jeweils zu diesen Klassen assoziierten, in Hyperkegeln lokalisierten  
Morphismen $\sigma_{\cK_1}$, $\tau_{\cK_2}$ komponiert werden können, 
$\tau_{\cK_2} \scirc \sigma_{\cK_1}$. Die resultierenden  
partiellen Zustände 
$\omega \scirc  \tau_{\cK_2} \scirc \sigma_{\cK_1} \rest \cA(V)$
gehören unabhängig von der Wahl von $\cK_1$, $\cK_2$
alle zu einer bestimmten  Ladungsklasse $L_3$, die sich ebenfalls 
durch lokalisierte Morphismen $\rho_{\cK_3}$
beschreiben lässt. Insbesondere gibt es also unitäre 
Operatoren $W_{3 2 1} \in \cA(V)$, die diese Morphismen verknüpfen, 
$\rho_{\cK_3} = \mbox{Ad} \, W_{3 2 1} \scirc \tau_{\cK_2} \scirc \sigma_{\cK_1}$.
Dies Resultat zeigt, dass bei der Komposition von Ladungen 
keine neuartigen Typen von Infrarotwolken entstehen, deren
Superauswahlsektoren sich in~$V$ unterscheiden lassen.

Bei unserer heuristischen Diskussion der Ladungserzeugung waren 
wir von der Existenz von 
entgegengesetzten Ladungen ausgegangen, die sich gegenseitig 
neutralisieren. Es zeigt sich, dass auch diese empirische Tatsache 
bereits in der Struktur der lokalisierten Morphismen kodiert 
ist~\cite{BuRo}: Zu jeder Ladungsklasse $L$ von Zuständen und 
gegebenem Lichtkegel $V$ mit dazu 
assoziierten in Hyperkegeln 
lokalisierten Morphismen $\sigma_\cK$ existiert eine 
konjugierte Ladungsklasse $\overline{L}$ mit 
entsprechend lokalisierten 
Morphismen $\overline{\sigma}_\cK$ für die gilt 
$\sigma_\cK \scirc \overline{\sigma}_\cK = 
\overline{\sigma}_\cK  \scirc \sigma_\cK  = \iota$. Das heißt,  
alle Effekte der Ladungen in $L$ lassen sich durch Komposition mit 
bestimmten Konfigurationen der konjugierten Ladungen aus 
$\overline{L}$ vollständig neutralisieren und umgekehrt.  
Die mit  $\overline{L}$ affiliierten Zustände werden daher 
als zu $L$ ladungskonjugierte Antimaterie interpretiert.

Eine weitere fundamentale Eigenschaft von Ladungen, 
nämlich die Tatsache, dass sie bei Vertauschung  in Zuständen 
(je nach Ladungstyp) entweder der Bose-- oder der
Fermistatistik genügen, lässt sich ebenfalls aus den 
Eigenschaften der lokalisierten Morphismen herleiten \cite{BuRo}. 
Dazu betrachtet man bei gegebener Ladungsklasse 
$L$ und Lichtkegel $V$ zwei assoziierte, beliebig lokalisierte Morphismen 
$\sigma_{\cK_1}$, $\sigma_{\cK_2}$ und 
vergleicht die resultierenden komponierten Morphismen 
$\sigma_{\cK_1} \scirc \sigma_{\cK_2}$ bzw.\ $\sigma_{\cK_2} \scirc \sigma_{\cK_1}$. 
Wie oben erläutert, gehören diese komponierten Morphismen 
zur gleichen Ladungsklasse (mit verdoppelter Ladung) und sind durch
unitäre Verknüpfungsoperatoren miteinander verbunden. 
Tatsächlich gibt es eine kanonische Wahl für diesen 
Verknüpfungsoperator, der in einem 
feldtheorischen Rahmen als gruppentheoretischer 
Kommutator von ladungstragenden 
(und daher nicht observablen) unitären Operatoren 
gedeutet werde könnte. 

Um diesen Sachverhalt zu beleuchten, nehmen wir für einen 
Moment an, dass es ladungstragende, in $\cK_j$ lokalisierte 
unitäre Operatoren $V_j$ gibt so dass $\sigma_{\cK_j} = \mbox{Ad} \, V_j$, 
$j = 1,2$. Der Operator $W_{2 1} = V_2 V_1^{-1}$ wäre dann 
ein Verknüpfungsoperator zwischen $\sigma_{\cK_1}$ und 
$\sigma_{\cK_2}$ im Sinne der oben angegebenen 
Definition und der
gruppentheoretische Kommutator $V_2 V_1 V_2^{-1} V_1^{-1}$
wäre ein Verknüpfungsoperator zwischen 
$\sigma_{\cK_1} \scirc \sigma_{\cK_2} = \mbox{Ad} \, V_1 V_2$
und $\sigma_{\cK_2} \scirc \sigma_{\cK_1} = \mbox{Ad} \, V_2 V_1$.
Der interessante Punkt ist nun, dass sich dieser 
gruppentheoretische Kommutator auch ohne Kenntnis von
ladungstragenden Operatoren berechnen lässt gemäß
$$V_2 V_1 V_2^{-1} V_1^{-1} = 
(V_2 V_1^{-1})   V_1 ( V_1  V_2^{-1}) V_1^{-1}  =
W_{2 1} \, \mbox{Ad} \, V_1 (W_{2 1}^{-1}) = 
W_{2 1} \sigma_{\cK_1}(W_{2 1}^{-1}) \, .
$$
Auf der rechten Seite dieser Gleichung steht eine Größe,
die in unserem nur auf Observablen aufgebauten   
Formalismus bestimmt werden kann.

Diese Überlegungen legen es nahe, für die gegebenen 
Morphismen $\sigma_{\cK_1}$, $\sigma_{\cK_2}$
und entsprechendem Verknüpfungsoperator $W_{2 1} \in \cA(V)$ 
den Operator 
$\varepsilon(\sigma_{\cK_1}, \sigma_{\cK_2}) = 
W_{2 1} \, \sigma_{\cK_1}(W_{2 1}^{-1})$ einzuführen. 
Man stellt fest, dass $\varepsilon(\sigma_{\cK_1}, \sigma_{\cK_2})$ nicht von
der Wahl von $W_{2 1}$ abhängt sowie 
die komponierten Morphismen 
$\sigma_{\cK_1} \scirc \sigma_{\cK_2}$ und $\sigma_{\cK_2} \scirc \sigma_{\cK_1}$
verknüpft. In diesem Sinne ist 
\begin{figure}[h] 
 \hspace*{63mm}
\epsfig{file=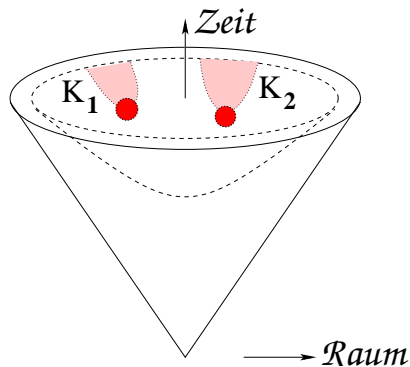,height=48mm}
\caption{Raumartig getrennte hyperbolische Kegel}
 \label{statistik}
 \end{figure}
der Operator kanonisch. Der physikalisch bedeutsame Punkt ist 
nun, dass für Morphismen, die in raumartig getrennten Hyperkegeln 
$\cK_1, \cK_2$ lokalisiert sind, siehe Fig.\ \ref{statistik}, der Operator 
$\varepsilon(\sigma_{\cK_1}, \sigma_{\cK_2})$
eine besonders einfache Form hat. Man kann nämlich 
zeigen \cite{BuRo}, dass es zu jeder Ladungsklasse~$L$ eine Zahl 
$\varepsilon_L \in \{\pm 1 \}$ (genannt Statistikparameter) gibt, so dass 
$\varepsilon(\sigma_{\cK_1}, \sigma_{\cK_2}) = \varepsilon_L 1$
für alle mit der Klasse assoziierten, in raumartig getrennten  
Hyperkegeln lokalisierten Morphismen $\sigma_{\cK_1},
\sigma_{\cK_2}$. Die physikalische 
Bedeutung dieses Ergebnisses wird klar, wenn man 
es durch die oben betrachteten  
ladungstragenden Operatoren $V_j$, $j=1,2$
reformuliert: Falls $\varepsilon_L = 1$ müssen diese 
Operatoren bei raumartigen Abständen 
miteinander kommutieren (bosonischer Fall), 
falls $\varepsilon_L = -1$ müssen sie antikommutieren
(fermionischer Fall). Jede Ladungsklasse $L$ genügt also  
entweder der Bose oder der Fermi Statistik. Andere 
Möglichkeiten gibt es für die hier betrachteten einfachen
Ladungen nicht. Man kann ferner zeigen \cite{BuRo}, dass die Statistik der 
konjugierten Klasse $\overline{L}$ stets mit der von 
$L$ übereinstimmt, $\varepsilon_{\overline{L}} = \varepsilon_L^{ }$. 

Es sei betont, dass diese Ergebnisse ganz wesentlich von 
der Tatsache abhängen, dass die physikalische Raumzeit
vierdimensional ist. In hypothetischen Welten niedrigerer 
Dimension würde die gleiche Analyse ergeben, dass 
Ladungsklassen auch einer anderen (z.B.\ anyonischen) Statistik   
genügen können. Man gelangt somit,  wie im massiven Fall \cite{Ha}, 
auch in Theorien mit masselosen Teilchen zu einem vertieften 
Verständnis der Wurzeln der Teilchenstatistik. 

\section{Kovarianz und Spektrum}

Ein weiterer physikalisch wichtiger Punkt ist die Frage, wie man im 
vorliegend Rahmen die relativistische Invarianz der Ladungsklassen und
die energetischen Eigenschaften der entsprechenden 
Zustände beschreibt. Wie bereits erwähnt, wirkt auf einen 
gegebenen Lichtkegel $V$ in natürlicher Weise nur 
die Semigruppe $\cS \subset \cP$ der Lorentztransformationen 
und zukunftsgerichteten Translationen, die $V$ auf bzw.\ in sich 
abbilden. Wir müssen daher zunächst diskutieren, wie man in physikalisch 
sinnvoller Weise den Transport von lokalisierten
Morphismen $\sigma$ \mbox{definiert}. (Da wir 
im folgenden einen festen Morphismus betrachten 
spielt dessen Lokalisationsgebiet keine Rolle.)

Zur Lösung dieses Problems gehen wir von einigen physikalisch
motivierten Annahmen aus. Für eine gegebene Transformation
$\lambda \in \cS$ sollte es im Prinzip möglich sein,  
die durch $\sigma$ beschriebene Ladungserzeugung in $V$ 
in exakt der gleichen Weise auf dem transformierten 
Lichtkegel \mbox{$\lambda V \subset V$} zu bewerkstelligen
(Wiederholbarkeit des Experiments). 
Es sollte also Morphismen ${}^\lambda \sigma$ auf $\cA(V)$ 
geben für die gilt ${}^\lambda \sigma \scirc \alpha_\lambda 
= \alpha_\lambda \scirc \sigma$, wobei daran erinnert sei,  
dass \mbox{$\alpha_\lambda \, \cA(V) = \cA(\lambda V)$}. 
Der Ladungsinhalt ändert sich beim
Transport nicht, die Morphismen ${}^\lambda \sigma$ 
gehören somit alle zur gleichen Ladungsklasse und es gibt unitäre
Verknüpfungsoperatoren $W_\lambda \in \cA(V)$, so dass 
$\mbox{Ad} \, W_\lambda \scirc {}^\lambda \sigma = \sigma$,
$\lambda \in \cS$. 
Die relativistische Invarianz der Theorie kommt dann dadurch zum 
Ausdruck, dass die durch  $W_\lambda$ beschriebenen 
physikalischen Operationen auf allen transportierten Lichtkegeln 
$\mu V$, $\mu \in \cS$ in analoger Weise operieren, d.h.
$\alpha_\mu(W_\lambda)$ ist auch Verknüpfungsoperator zwischen
den transportierten Morphismen 
${}^{\mu \lambda} \sigma$ und ${}^{\mu} \sigma$ für
$\lambda, \mu \in \cS$. (Hier benutzt man die Tatsache 
dass $\cS$ eine Semigruppe ist, dass also das dem
sukzessiven Transport entsprechende Produkt erklärt ist.)  
Wir fassen diese physikalisch erwarteten Eigenschaften in 
folgender Definition zusammen. 

\vspace*{4mm} 
\noindent \textbf{\textit{Kovariante Morphismen:}} \ 
Ein Morphismus $\sigma$ auf $\cA(V)$ heißt kovariant, wenn es 
eine dazu assoziierte Familie von transportierten Morphismen 
${}^\lambda \sigma$, $\lambda \in \cS$ gibt mit den oben angegebenen 
Eigenschaften. 

\vspace*{2mm} 
Es ist bemerkenswert, dass diese zwar physikalisch gut begründete 
aber implizite Charakterisierung der kovarianten Morphismen ausreicht,  
um die gewohnte Formulierung der relativistischen
Kovarianz zu etablieren \cite{BuRo}. 
Ausgehend von der durch das Vakuum fixierten unitären Darstellung 
$U$  der vollen Poincar\'e Gruppe $\cP$ kann man nämlich  
zeigen, dass die gemäß $U_\sigma(\lambda) = U(\lambda) W_\lambda$, 
$\lambda \in \cS$ definierten unitären Operatoren die 
Transformation der Observablen in der Ladungsklasse von 
$\sigma$ implementieren, 
$$
\mbox{Ad} \, U_\sigma(\lambda) \scirc \sigma = \sigma \scirc
\alpha_\lambda \, , \quad \lambda \in \cS \, .
$$
Darüber hinaus existiert stets eine Wahl der Verknüpfungsoperatoren
$W_\lambda$, so dass sich die Operatoren $U_\sigma(\lambda)$, $\lambda \in \cS$ 
eindeutig zu einer unitären Darstellung von $\cP$ 
(bzw.\ der Überlagerungsgruppe von $\cP$) fortsetzen lassen. 
Beobachter in $V$ können somit in eindeutiger
Weise über den Energieinhalt und Drehimpuls (Spin) von Zuständen in 
der gegebenen Ladungsklasse sprechen. Diese ergeben sich aus den Generatoren
der Darstellung $U_\sigma$, die jedoch nicht direkt als 
Observablen \mbox{gedeutet} werden können. Observablen im üblichen 
Sinne sind die unitären (und damit im Sinne der 
Spektraltheorie normalen) Verknüpfungsoperatoren 
$W_\lambda = U(\lambda)^{-1} U_\sigma(\lambda) \in \cA(V)$, 
$\lambda \in \cS$ aus denen man z.B.\ Informationen über den 
Energieunterschied zwischen Zuständen in der Ladungsklasse
des Vakuums und der geladenen Klasse erhält. Mit der
getroffenen Wahl des Vakuums als Referenzzustand und
der sich daraus eindeutig ergebenden Darstellung $U$ von 
$\cP$ kann man diese Informationen dann mithilfe der 
Darstellungen $U_\sigma$ in gewohnter Weise beschreiben. 
 
Ein weiteres Resultat, das die physikalische Konsistenz 
unseres theoretischen Zugangs untermauert, betrifft die  
Form des Energie--Impuls Spektrums in den Ladungsklassen. 
Durch eine detaillierte Analyse des sich 
bei der Komposition von konjugierten Ladungsklassen ergebenden 
Spektrums lässt sich zeigen \cite{BuRo}, dass die Generatoren 
der raumzeitlichen Verschiebungen $U_\sigma \rest \RR^4$
in allen Ladungsklassen die relativistische Spektrumsbedingung 
erfüllen (Positivität der Energie in allen Inertialsystemen).
Diese energetische Stabilität der Ladungsklassen ist also 
eine Konsequenz der Struktur der Observablen und des 
Vakuums und muss nicht extra postuliert werden.  

\section{Fazit}

Die vorangegangene Diskussion hat verdeutlicht, dass 
die notorischen Infrarotprobleme bei der Interpretation 
von Quantenfeldtheorien mit langreichweitigen Kräften  
auf einer Überidealisierung beruhen:
Nimmt man im Rahmen der Theorie wie gewöhnlich an, 
dass der masselose Teilcheninhalt der Zustände durch 
Messungen im gesamten Minkowskiraum bestimmt werden kann,
so ist man gezwungen, eine Vielzahl unendlicher 
Infrarotwolken masseloser Teilchen in Betracht zu ziehen. 
Berücksichtigt man jedoch die Existenz des Zeitpfeils, 
so ist offensichtlich, dass sich Messungen bestenfalls
in zukunftsgerichteten Lichtkegeln durchführen lassen.
Versäumte Messungen in der Vergangenheit bedeuten 
im Allgemeinen einen unwiederbringlichen Verlust an Informationen, 
und die Theorie kann sich daher auf die 
Beschreibung und Interpretation von Lichtkegeldaten
beschränken. 

Wie wir gesehen haben, führt diese Beschränkung  
zu einer sowohl mit den Symmetrien des Minkowskiraums 
als auch dessen kausaler Struktur verträglichen 
Infrarotregularisierung. Das aus der Einsteinschen
Kausalitätsforderung 
ableitbare Huygenschen Prinzips bewirkt die geometrische 
Aufteilung des globalen masselosen Teilcheninhalts
in einen in Lichtkegeln experimentell zugänglichen essentiellen und 
einen in der Vergangenheit abgestrahlten und daher 
für alle Zeiten verlorenen marginalen Anteil. 
Angesichts dieser Fakten wird im Rahmen 
der Theorie verständlich, weshalb sich globale 
Infrarotwolken nicht scharf unterscheiden lassen. 

Auch die Frage, welche physikalischen Aspekte einer Theorie 
in Lichtkegeln bestimmt werden können,  
konnte beantwortet werden. Dies sind zum einen 
die Eigenschaften massiver Teilchen, insbesondere deren 
Ladungen und Statistik, sowie die Existenz und Eigenschaften 
von Antimaterie. Darüber hinaus lassen sich aus den   
auf Lichtkegel eingeschränkten partiellen Zuständen 
auch deren energetische und  
rotatorischen Eigenschaften bestimmen. Materie kann also   
auch in diesem neuen theoretischen Zugang 
detailliert beschrieben werden. 

Der in Lichtkegeln nachweisbare masselose Teilcheninhalt lässt sich
durch Zustände mit endlicher Teilchenzahl (also im Fockraum) beschreiben.
Es sei betont, dass dies auch masselose Teilchen einschließt, die 
den Lichtkegel aus fernen Galaxien erreichen. Diese Teilchen wurden
zwar vor Äonen erzeugt, doch kommen sie nicht aus der 
Vergangenheit des Beobachters, sondern  aus zum 
Beobachter zuvor raumartig getrennt liegenden Gebieten.
Stellt man sich nun vor, dass durch diese Teilchen Informationen
über das Ergebnis von Experimenten unserer 
\mbox{extragalaktischen} 
Kollegen übermittelt werden, so kann man sich die Frage stellen,
ob diese Informationen im Widerspruch zu den in hiesigen Lichtkegeln 
gewonnenen physikalischen Einsichten stehen könnten. 
Die Antwort ist beruhigend: Auch die extragalaktischen Kollegen
können ihre Experimente bestenfalls in Lichtkegeln 
durchführen und man kann 
zeigen \cite{BuRo}, dass die Vorhersagen der Theorie nicht 
von der Wahl eines Lichtkegels abhängen. Somit ist die innere 
Konsistenz unserer zwar ungewohnten aber dafür problemfreien 
Interpretation der relativistischen Quantenfeldtheorie gesichert.

\bigskip 
\section*{Danksagung} 
Mein Dank gilt dem Evangelischen Studienwerk, 
besonders Herrn Prof.\ Dr.\ Eberhard Müller,  
für die langjährige fruchtbare Zusammenarbeit und 
stete Förderung des Promotionsschwerpunktes ``Wechselwirkung''.

\end{document}